\documentclass[12pt,preprint]{aastex}

\newcommand{\siml}{\lower4pt \hbox{$\buildrel < \over \sim$}}
\newcommand{\simg}{\lower4pt \hbox{$\buildrel > \over \sim$}}

\begin{document}
\title{A model for the challenging ``bi-drifting'' phenomenon
in PSR J0815+09}
\author{G. J. Qiao$^{1}$, K. J. Lee$^{1}$, B. Zhang$^{3}$. R. X. Xu$^1$,
\& H. G. Wang$^{2}$}

\input psfig.sty

\affil{ $^1$Department of Astronomy, Peking University,
Beijing 100871, China, gjn@pku.edu.cn \\
$^2$ Center for Astrophysics, Guangzhou University, Guangzhou
510400, China, cosmic008@263.net\\
$^3$ Department of Physics, University of Nevada, Las Vegas,
Nevada 89154-4001, bzhang@physics.unlv.edu}

\begin{abstract}

A new drifting pulsar, PSR J0815+09, was discovered in the Arecibo
drift-scan searches \citep{Mc04}. An intriguing feature of this
source is that within the four pulse components in the integrated
pulse profile, the sub-pulse drifting direction in the two leading
components is opposite from that in the two trailing components.
In view that the leading theoretical model \citep{RS75} for pulsar
sub-pulse drifting can only interpret one-direction sub-pulse
drifting, the observed \emph{bi-drifting} phenomenon from PSR
J0815+09 poses a great challenge to the pulsar theory. The inner
annular gap (IAG), a new type of inner particle accelerator, was
recently proposed to explain both $\gamma$-ray and radio emission from
pulsars \citep{QLWXH04}. Here we show that the coexistence of the IAG
and the conventional inner core gap (ICG) offers a natural
interpretation to the bi-drifting phenomenon. In particular, the
peculiar drifting behavior in PSR J0815+09 can be reproduced
within the inverse Compton scattering (ICS) model for pulsar radio
emission.

\end{abstract}


\section{Introduction}

The sub-pulse drifting phenomenon observed in many pulsars has been
widely regarded as a powerful tool to probe the pulsar inner
magnetospheric structure and radiation mechanism (Ruderman \&
Sutherland 1975, hereafter RS; Gil, Melikidze \& Geppert 2003).
It may be linked to the
physical properties of the surface of pulsars (RS75), shedding
light on the nature of pulsars, e.g. whether they are normal neutron
stars or bare strange stars (Xu, Qiao \& Zhang 1999).

In the classical drifting model, sub-pulse patterns are
manifestations of some sparks passing along a ring (``carousel'')
in the polar cap region circulating the magnetic axis (RS75). This
interpretation is supported by detailed analysis of drifting data
\citep{DR99,DR01,AD01,GS03}. The RS model is the leading model to
interpret the observations quantitatively. More detailed analysis
revealed that the circulation speed in a pure vacuum gap is too
high when compared with the observations \citep{GGMG03}. Such a
discrepancy may be resolved through introducing partial screening
of the gap electric field so that the inner gap is not purely
vacuum \citep{GGMG03,CR80,UM95}. These models are successuful to
interpret many observations, including time variations of the drift
rate and changes of the apparent drift direction \citep{GGMG03}.

Recently, a new drifting pulsar, PSR J0815+09, was discovered in
the Arecibo drift-scan searches \citep{Mc04}. An intriguing
feature of this source is that within the four pulse components in
the integrated pulse profile, the sub-pulse drifting direction in
the two leading components is opposite from the one in the two
trailing components. We call this ``bi-direction drifting'' or
``bi-drifting'' phenomenon. This phenomenon poses a great
challenge to the RS model and its variants, since in all these
models, the drifting direction in different spark-rings is expected to
be the same. Recently, Qiao et al. (2004) suggested that an inner
annular gap (IAG) likely coexists with the conventional inner core
gap (ICG) in pulsars (especially when pulsars are bare strange
stars), and proposed a phenomenological model to illustrate the
origin of the $\gamma$-ray and radio radiation from pulsars. In
this paper, we propose that the bi-drifting phenomenon is a
natural consequence of the coexistence of the IAG and the ICG.
Furthermore, the complicated drifting patterns in PSR J0815+09
could be reproduced if the pulsar radio emission is dominantly
generated through coherent inverse Compton scattering (ICS)
processes \citep{QL98,XLHQ00,QLZH01}.

In \S 2, we introduce both the ICG and the IAG in pulsars, and the ICS
model for radio emission is introduced in \S 3. Numerical simulations
of the observed bi-drifting phenomenon is presented in \S 4, and the
conclusions and discussions are presented in \S 5.

\section{The inner core gap (ICG) and the inner annular gap (IAG)}

\begin{figure}
\plotone{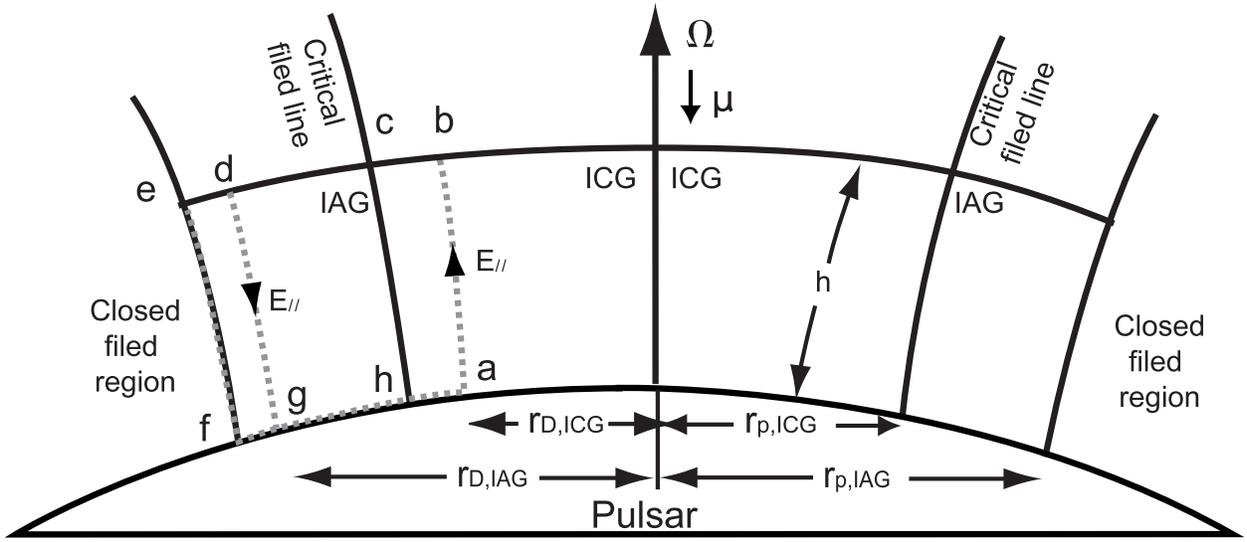} \caption{The inner annular gap (IAG) and the
inner core gap (ICG). $r_{\rm p, IAG}$ and $r_{\rm p, ICG}$ are
the radii of the IAG and the ICG, respectively. The two
gaps are divided by the boundary line 'ch', which is the critical
magnetic line that passes through the intersection of the null-charge
line and the light cylinder. The dashed lines are used to
illustrate the electric property of the gaps discussed in section
2. The parameters used in our simulations are presented in the text.
}
\label{cap}
\end{figure}

The pulsar's polar region defined by the last open magnetic field
lines actually
includes two parts separated by the \emph{critical field
lines} (RS75, Qiao et al. 2004). Oppositely-charged particles
leave the two region respectively, (e.g. the positive charges leave
the central region while the negative charges leave the annular region
for a pulsar with ${\rm \Omega \cdot B<0}$). The deviation of the
charge density from the Goldreich-Julian (1969, hereafter GJ)
charge density has opposite signs in the two regions.
Qiao et al. (2004) suggested that in principle there could form
two kinds of sparking inner gaps, i.e. the
conventional inner core gap (ICG) above the central part of the
polar region and an inner annular gap (IAG) above the annular part of
the polar region. Both gaps can be high energy particle accelerators.
A geometric model involving
emission from both gaps can interpret the diverse morphology
of both gamma-ray and radio emission from gamma-ray pulsars. The
geometric structure of the two kinds of polar gap is plotted in
Fig.\ref{cap}.

For the inner gaps, the typical gap height is
smaller than the polar cap radius, and one can use the 1-D
approximation of the Poisson equation in the co-rotate frame,
i.e. $\partial E_{\parallel}/\partial
x=4\pi(\rho-\rho_{\rm GJ})$, where $x$ is the longitudinal
distance measured from the surface along the curved magnetic field
lines, and $E_{\parallel}$ is the parallel component of the
electric field with respect to the magnetic field, $\rho$ is the charge
density and $\rho_{\rm GJ}$ is the GJ charge density. This
equation governs the electric field along the magnetic field
(i.e. segments ``ab'', ``ch'', ``dg'' and ``ef'' in Fig.\ref{cap}).
The boundary condition equation is $E_{\parallel}=0$ at the upper
boundary of the gap (i.e. segment ``edcb'').

When the gaps are re-generated after each sparking process, since
opposite charged particles leave the ICG and the IAG regions,
respectively, the sign of $\rho-\rho_{\rm GJ}$ is the opposite in
the two gaps. The Poisson equation above shows that the directions
of $E_{\parallel}$ are also different in the IAG and ICG.
We expect that at the boundary between ICG and IAG (line ``ch'')
the parallel electric field vanishes. That is $\rm \int_{c}^{h}
E\cdot ds=0$.

We handle the electrodynamics in the co-rotate frame. For any
close circuits (e.g. ``defg'', ``abch''), one has $\oint E\cdot
ds=0$. In the closed magnetic field region ${\bf E \cdot B}=0$, so
$\int_{e}^{f}{\bf E\cdot ds}=0$. As discussed in RS75, we have
$\int_{f}^{g}{\bf E\cdot ds}=0$. Thus for the IAG, one has
$\int_{g}^{d} {\bf E\cdot ds}+\int_{d}^{e} {\bf E\cdot ds}=0$.
When considering the boundary condition between the two gaps one has
$\int_{c}^{h}{\bf E\cdot ds}=0$. Also $\int_{h}^{a}{\bf E\cdot
ds}=0$ is satisfied for the same reason as in the discussion of the IAG
case. One can get $\int_{a}^{b} {\bf E\cdot ds}+\int_{b}^{c}
{\bf E\cdot ds}=0$. Therefore for the IAG and the ICG the
perpendicular electric field is directly linked to the parallel
electric field.
Becasue the parallel electric fields in the IAG and the ICG have
different directions (i.e. $\int_{a}^{b} {\bf E\cdot ds}$ and
$\int_{g}^{d} {\bf E\cdot ds}$ have different signs), the
perpendicular
electric fields in the two gaps are also different. The drifting
velocity ${\bf v=E \times B/|B|^2}$ have opposite signs in the ICG
and the IAG, because the direction of magnetic fields is the same
in the two gaps. This proposes a fundamental physical process to
understand the bi-drifting phenomenon.

\section{The inverse Compton scattering (ICS) model of pulsar radio
emission}

Any radiation model possessing a symmetrical radiation beam
is not applicable to PSR J0815+09. If the radiation beam
is symmetrical, the 2nd and 3rd components should come from the same
beam, so they should have a same drifting sense which contradicts the
data. When the aberration and retardation effects are considered, the
ICS (inverse Compton scattering) model naturally predicts an asymmetric
beam and offers a natural mechanism to interpret the phenomenon.

The ICS model \citep{QL98,XLHQ00,QLZH01,QLWXH04} suggests that the
observed radio emission comes from different emission heights and each
sparking ring can, in principle, lead to several emission cone-shaped
beams.  Following Qiao and Lin (1998), one can get the so-called
``beam-frequency figure'' of the ICS model, which naturally gives rise
to a narrow central core emission component and two conal components,
meeting Rankin's (1983; 1993) empirical proposal from radio pulsar
data. Another remark is that the three-component scheme is an average
picture, which resembles Manchester's (1995) ``window function''
scheme. Depending on the line of sight, different number of emission
components at a particular observing frequency can be observed. That
PSR J0815+09 is observed as a drifting pulsar means that the line
of sight sweeps across the beam rim rather than cutting the rim
tangentially. For the latter case periodical variation of the
pulse intensity rather than sub-pulse drifting should be observed. It
is likely that the line of sight misses the narrow core beam and only
sweeps the two conal beams. In this way, there are four conal
radiation beams in the system, two of which come form the IAG while the
other two come from the ICG. Generally the four beams should form
eight pulse components. There are still several important issues for
determining the beam morphology. One thing to put in consideration is
the aberration and retardation effects.  Another one is the radiation
process of inverse Compton scattering.

Because different emission components are emitted at different
heights, the retardation and aberration effects must be taken into
account, and they are found to be important to determine the real
emission morphology. These two effects will smear the leading
radiation components and strengthen the trailing components (Qiao \&
Lin 1998, Qiao et al. 2004, Dyks et al. 2004). From the simulations
(Qiao \& Lin 1998, Dyks et al. 2004), it is found that the trailing
components are about 2 orders of magnitude brighter than the leading
components if the radio waves are superposed coherently.
For incoherent superposition, the contrast should be still 1 order of
magnitude.

In the ICS model, the observed radiation comes from the inverse Compton
scattering between the high energy secondary particles and the low
frequency electromagnetic wave (EM wave) generated by the polar gap
sparking process. The intensity of the low frequency EM wave is a
function of the direction angle $\chi$ (i.e. $\propto \sin^{2}\chi$),
where $\chi$ is the angle between the radiation direction and the
electric dipole moment. When taking the rotation effect into account,
the leading and the trailing radiation sites at a same height do not
have the same angle respected to the direction of the spark's electric
moment, so that they do not possess the same intensity of the low
frequency EM wave. This effect will intensify the leading components
and weaken the trailing components. The ratio of the leading low
frequency wave intensity ($I_{\rm L}$) to trailing intensity ($I_{\rm
T}$) can be given by the equation $I_{\rm L}/I_{\rm
T}=\sin^{2}(\theta+\sin^{2}\theta)/\sin^{2}(\theta-\sin^{2}\theta)$,
when the inclination angle is $90^{o}$ and the spark is a pure
dipole.  This effect can lead to an intensity contrast of 2
orders of magnitude between the leading and trailing pulse components
for coherent radiation or of 1 order of magnitude for incoherent
radiation. Beside the two effects above, other potential
mechanisms may also play a role. For example, if the polar gap
sparks are triggered by the incoming $\gamma$-ray photons, the geometry
configuration of the pulsar, the beam direction and the $\gamma$-ray
photon source direction respected to the pulsar will all affect the
intensity ratio of the leading and the trailing components.

Comparing among these effects, the first one is the common situation.
So in most cases, the trailing components may be the ones that are
observed. However, we can not exclude other possibilities discussed
above, and will take it as a basic assumption that the radiation beam
is asymmetrical, and that only one half of the components can be
observed. Such an assumption has received observational
supports, since some pulsars are already observed to have asymmetrical
pulse profiles. This assumption can be tested by high quality
polarization observations, which may indicate whether the leading or
the trailing components are observed.

\section{Simulations of sub-pulse drifting}

To simulate the drifting pulse patterns, two parameters are included,
i.e. the number of the sub-beams in each beam and the drifting rate of
each pulse component. The numbers of the sub-beams are calculated
theoretically, and the drifting rates can be obtained from the
simulations, which would be used to infer the dynamical
structure of both gaps.

Gil \& Sendyk (2000) suggests that
the number of the sparks that can be observed
in drifting pattern is given by $N \simeq 2 \pi r_{\rm D}/h$,
where $N$ is the number of sparks, $h$ is the height of the
gap, $r_{\rm D}$ is the radius of the sub-pulse drifting trajactory
which is given by $r_{\rm D}=r_{\rm p}-h/2$. For the parameters of PSR
J0815+09 (P=0.645s and $B=3\times10^{11}$G, McLaughlin et al.
2004), it can be shown that $r_{\rm p,\rm IAG}=R^{3/2} R_{\rm
lc}^{-1/2}\simeq (1.45 \times 10^4 {\rm cm}) p^{-1/2}\simeq 1.80 \times
10^4 $ cm for the IAG and $r_{\rm p, ICG} =(2/3)^{3/4}R^{3/2}R_{\rm
lc}^{-1/2}\simeq(1.07 \times 10^4 {\rm cm} )p^{-1/2}\simeq 1.35 \times
10^4$ cm for the ICG, where $R$ is the stellar radius, $R_{\rm lc}$ is
the radius of light cylinder. For a resonance inverse Compton
scattering gap, the gap height is $h\simeq (1.1\times 10^3 {\rm cm})
B_{12}^{-1}p^{1/3} \simeq 3.2\times 10^3 $ cm \citep{ZQH97}, where
$B_{12}$ is the surface magnetic field in unit of $10^{12}$G, where
the curvature radius of the magnetic fields is taken as $10^6 $cm. So
the IAG holds $N \simeq 33$ observed sparks while the ICG holds $ N
\simeq 21$ sparks. These two parameters will be put in our
simulation for the drifting patterns. Assuming a same intensity for
each component within the gap, we can also calculate
the integrated pulse profile.

\begin{figure}
\plotone{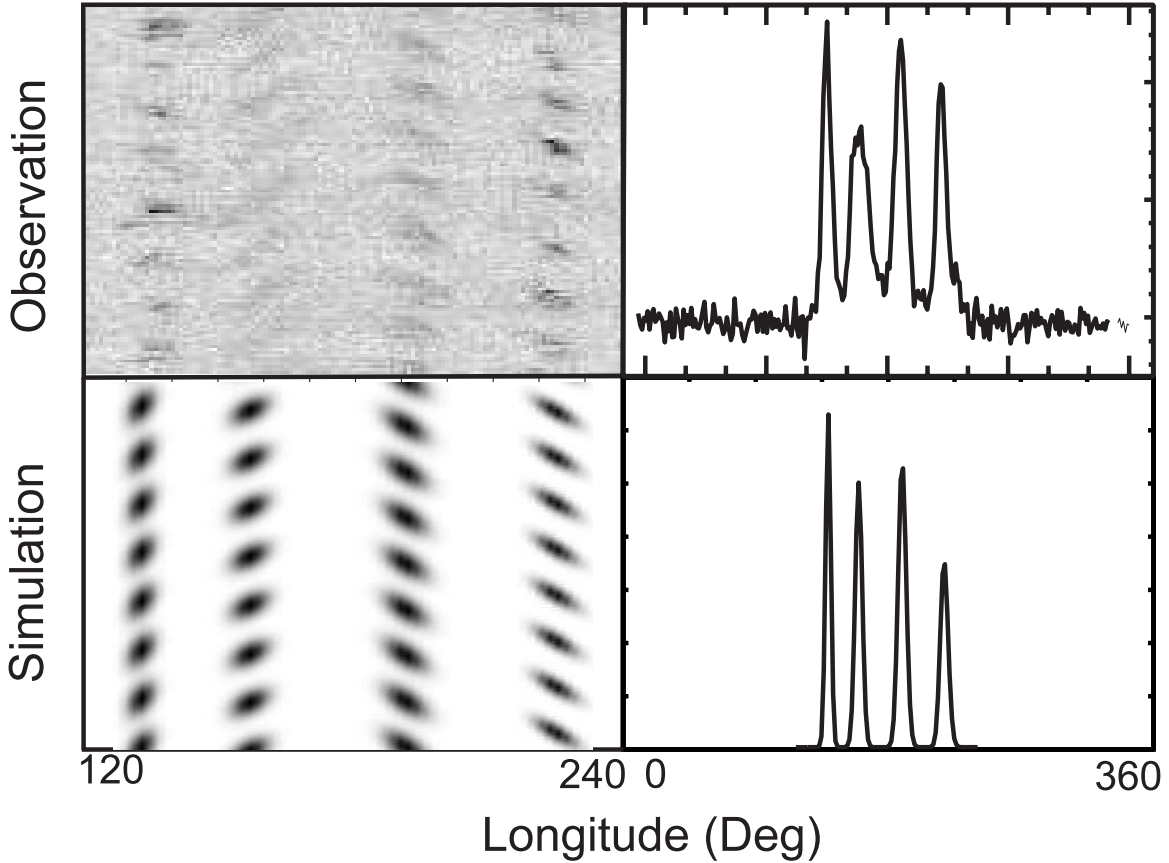}
\caption{The observed sub-pulse drifting patterns and pulse profiles
as compared with our
simulations. The left two panels are the patterns of
drifting sub-pulses. The right two panels are the integrated pulse
profiles. The observational data are derived from McLaughlin et
al. 2004. In the simulation, the ICG has 21 sparks circularly drifting
around the magnetic axis with a period of 210s, while the IAG has 33
sparks with a drifting period of 310s. The spark number is given by
theoretical estimates and the drifting periods are inferred through
matching the simulation with the observations.
Lacking of polarization data, we can not constrain the radiation
geometry. In the simulation, the radiation geometry are choosen
relatively arbitrarily. The angular radii for the four beams are 12,
28, 46 and 63 degrees, the inclination angle is 45 degrees, and the
impact angle is 5 degrees.
}
\label{drift}
\end{figure}

The drifting rates of the sparks depend on the electric structure of
the gaps. However many uncertainties involved in the gaps
\citep{ZQLH97,GGMG03,QLWXH04} prevent us from retrieving solid
information for the drifting rates. Here we use the observational
drifting rates to constrain the gap models. From the theoretical
sub-beam number and the observed drifting patterns, we get the drifting
rates in both gaps. The simulated results and observations are
compared in Fig.\ref{drift}.

\section{Results and discussion}
By considering both the ICS process and the sparks from the IAG
and the ICG, we have demonstrated that the bi-drifting phenomenon
observed in PSR J0815+09 is naturally interpreted. This result also
lends support to the existence of the IAG. Some parameters
and effects related to our simulation are discussed as follows.

(1) The gap heights. Assuming that the spark diameter in the polar
gap is the same as the gap height \citep{GS00}, we have used the
resonance inverse Compton scattering induced polar gap model
\citep{ZQLH97} to calculate the gap height to get the spark number in
the polar cap. One requirement to the height of the sparks is that
some sparks must take place in the IAG. The height of a resonance
inverse Compton scattering induced gap in our simulation is consistent
with this condition.

(2) The drifting rates. The drifting rates derived from the
observation give us insights into the electric structure of the two
gaps. The average parallel electric field $E_{\parallel}$ in the gap
can be estimated with $E_{\parallel}\sim 4\pi h(\rho-\rho_{\rm GJ})$.
From \S 2, we know that the perpendicular electric field
$E_{\perp}$ can be estimated as $E_{\rm \perp}\simeq
E_{\parallel}h/\Delta r$ (see also Gil et al. 2003), where
$\Delta r$ is the transverse spatial dimension of the sparks and
has $\Delta r\simeq h/2\simeq1.6 \times 10^3 $ cm. The drifting
velocity of the sparks is therefore $v_{\rm D}=E_{\perp} c/B$,
where $c$ is the speed of light, and the period for a spark to make a
circle in both gaps can be written as
$\hat{P_{3}}=2\pi r_{\rm D}/v_{\rm D}$, or
\begin{equation}
\hat{P_{3}}\simeq \frac{B r_{\rm D} \Delta r} {2ch^{2}(\rho-\rho_{\rm GJ})}
\label{P3}
\end{equation}
The period for a subpulse to reappear at a same phase is defined by
$P_3 = \hat P_3/n$, where $n \propto r_{\rm D}$ is the number of
sparks in a circle.
Since both the IAG and the ICG have the same height $h$ and hence, the
same $\Delta r$, from Eq.(\ref{P3}) one gets
\begin{equation}
\frac{P_{\rm 3,ICG}}{P_{\rm 3,IAG}} \simeq \frac { r_{\rm \rm D,
IAG} \hat{P_{\rm 3}}_{\rm , ICG} } { r_{\rm D, ICG} \hat{P_{\rm
3}}_{\rm , IAG} } \simeq \frac{(\rho-\rho_{\rm GJ})_{\rm
IAG}}{(\rho-\rho_{\rm GJ})_{\rm ICG}} =
 \frac{E_{\parallel, \rm
IAG}}{E_{\parallel, \rm ICG}}, \label{ele}
\end{equation}

The simulation gives $\hat{P_{\rm 3,}} _{\rm IAG}/\hat{P_{\rm
3,}}_{\rm ICG}\simeq1.5$, and from \S 4, we have $r_{\rm D,
IAG}/r_{\rm D, ICG}\simeq1.4$ (Fig.~\ref{cap}). So
$E_{\parallel, \rm IAG}/E_{\parallel, \rm ICG}=P_{3,\rm IAG} /P_{3,\rm
ICG} \simeq0.9$.
This indicates that the absolute charge density deviation from
$\rho_{\rm GJ}$ in the ICG is roughly the same as that in the IAG,
although a charge deficit presents in the ICG while a charge excess
presents in the IAG.  Also the $E_{\parallel}$ in the ICG is roughly
the same as in the IAG but with a different sign. The sub-pulse
drifting phenomenon offers a diagnostic tool for the plasma near the
polar cap region.

The theoretical drifting period of a vacuum gap for PSR J0815+09
should be order of 1 s (Eq. (\ref{P3}) and put $\rho=0$) and is
much shorter than the fitted value $\sim10^2$ s. This has been
also noticed earlier by Gil et al. (2003). There are three
possible ways to solve this problem. One is to conclude that both
the IAG and the ICG are not vacuum gaps, so that only a small
charge density deviation from the GJ charge density is allowed. In
order to match the observations, only a $\siml 1\%$ deviation from
the GJ density in both gaps is required. This would naturally
reduce the drifting rates, and is consistent with the partial
screening picture (Gil et al. 2003). It is also consistent with
the XMM-Newton observation results for another drifting pulsar PSR
0943+10 (B. Zhang, D. Sanwal \& G. G. Pavlov 2004, in
preparation). In such a case, an unsteady space charge limited
flow may play an important role, and a detailed model is called
for. The second reason is overestimating the drifting
velocity, because the method above just give the average value of
$E_{\perp} $ which is larger than that at sparking location. The
third way to solve the problem is to take into account the
drifting dynamics, in which the drifting velocity is not the
$E\times B$ velocity. Again more detailed investigations are
needed. It should be noted that the subpulse drifting in the two
gaps depends on the global electric property of the pulsar. Two
gaps interact with each other via the boundary condition. So a
physically reasonable and complete drifting subpulse model should
at least includes a global electric solution of pulsars.

It is found that within the observation data span, the four pulse
components keep a phase relationship at least within 120 pulses
(77s). Our Eq.(2) suggests that $P_3$ in both gaps are roughly the
same. A further physical possibility is that the two gaps may interact
with each other. Dynamical system theories prove that for two
quasi-periodical systems, if there are some small nonlinear
interactions, the two systems will be locked into a nearby
frequency with a rational ratio \citep{JBB83}.  Because of the
interactions between the two gaps, the two drifting
frequencies may be locked into a relative constant ratio in our
model, so that the apparent $P_3$ could appear the same for all four
pulse components. Detailed interaction dynamics is needed to further
address this problem. Some other kinds of interaction between the two
gaps are also possible (e.g. Young 2004).

(3) High quality polarization observations are needed to finally
verify whether the observed components are the trailing or the leading
components.

In summary, the so called \emph{bi-drifting} phenomenon is a newly
observed touch-stone to test the radiation theories and the surface
physical properties of the pulsar. Our simulated results support the
coexistence of IAG and ICG. Further investigations are needed to
address the questions such as how such gaps are generated.

\begin{acknowledgements}
We are very grateful to the referee's insightful comments, which led
us to improve some of the very important points raised in this
paper. We also thank  Drs. Esamdin Ali \& Han, J. L. for their valuable
discussions. This work is supported by NSF of China (10373002,
10273001). B.Z. acknowledges NASA NNG04GD51G for support.

\end{acknowledgements}

\end{document}